\documentclass[journal]{IEEEtran}

\usepackage{xcolor,soul,framed} 

\colorlet{shadecolor}{yellow}
\usepackage[pdftex]{graphicx}
\graphicspath{{../pdf/}{../jpeg/}}
\DeclareGraphicsExtensions{.pdf,.jpeg,.png}

\usepackage[cmex10]{amsmath}
\usepackage{array}
\usepackage{mdwmath}
\usepackage{mdwtab}
\usepackage{eqparbox}
\usepackage{url}

\usepackage{amsmath}   
\usepackage{graphicx}  
\usepackage{amssymb}   

\usepackage{cite}
\usepackage[utf8]{inputenc}
\usepackage{amsmath}
\usepackage{graphicx}
\usepackage{textcomp}
\usepackage{xcolor}
\usepackage{float}
\usepackage[hidelinks]{hyperref}
\usepackage{booktabs}
\usepackage[numbers,sort&compress]{natbib}
\usepackage{algorithm}
\usepackage{algpseudocode}

\begin{document}
\bstctlcite{IEEEexample:BSTcontrol}
    \title{Agentic-AI based Mathematical Framework for Commercialization of Energy Resilience in Electrical Distribution System Planning and Operation}
    \author{
    Aniket~Johri, 
    Divyanshi~Dwivedi, 
    and Mayukha~Pal*
    \thanks{(*Corresponding author: Mayukha Pal)}
    \thanks{Mr. Aniket Johri is a Data Science Research Intern at ABB Ability
Innovation Center, Hyderabad 500084, India, and also an undergraduate at
the Department of Computer Science and Engineering, Indian Institute of
Technology Tirupati, Andhra Pradesh 517619, IN.}
    \thanks{Dr. Divyanshi Dwivedi is with ABB Ability Innovation Center, Hyderabad 500084, IN, working as Senior R\&D Engineer.}
    \thanks{Dr. Mayukha Pal is with ABB Ability Innovation Center, Hyderabad 500084, IN, working as Global R\&D Leader – Cloud \& Advanced Analytics
    (e-mail: mayukha.pal@in.abb.com).}
}
\maketitle

\begin{abstract}
The increasing vulnerability of electrical distribution systems to extreme weather events and cyber threats necessitates the development of economically viable frameworks for resilience enhancement. While existing approaches focus primarily on technical resilience metrics and enhancement strategies, there remains a significant gap in establishing market-driven mechanisms that can effectively commercialize resilience features while optimizing their deployment through intelligent decision-making. Moreover, traditional optimization approaches for distribution network reconfiguration often fail to dynamically adapt to both normal and emergency conditions. This paper introduces a novel framework integrating dual-agent Proximal Policy Optimization (PPO) with market-based mechanisms, achieving an average resilience score of 0.85 ± 0.08 over 10 test episodes. The proposed architecture leverages a dual-agent PPO scheme, where a strategic agent selects optimal DER-driven switching configurations, while a tactical agent fine-tunes individual switch states and grid preferences under budget and weather constraints. These agents interact within a custom-built dynamic simulation environment that models stochastic calamity events, budget limits, and resilience-cost trade-offs. A comprehensive reward function is designed that balances resilience enhancement objectives with market profitability (with up to 200x reward incentives, resulting in 85\% of actions during calamity steps selecting configurations with $\geq4$ DERs), incorporating factors such as load recovery speed, system robustness, and customer satisfaction. Over 10 test episodes, the framework achieved a benefit-cost ratio of 0.12 ± 0.01, demonstrating sustainable market incentives for resilience investment. This framework creates sustainable market incentives, with resilience prices guiding efficient resource allocation, transforming resilience into a revenue-generating utility service. 
\end{abstract}


%
\IEEEpeerreviewmaketitle


\section{Introduction}

The modern electrical power system faces unprecedented challenges from both the increasing penetration of renewable energy resources and the growing frequency of extreme weather events. These dual pressures have fundamentally shifted the focus of power system planning and operation from traditional reliability metrics to comprehensive resilience strategies that can withstand, adapt to, and rapidly recover from High-Impact, Low-Probability (HILP) events. As electrical distribution systems (EDS) are required to maintain uninterrupted power supply for critical loads including industrial production, healthcare facilities, national security infrastructure, and essential services, the economic implications of power disruptions have reached alarming proportions\cite{DIAHOVCHENKO2021107327}.

Major catastrophic events over the past decade have starkly illustrated the vulnerabilities inherent in existing power infrastructure. Hurricane Sandy in 2012 left 4.2 million customers without power for up to 10 days, while Hurricane Maria in 2017 affected 3.6 million residents in Puerto Rico \cite{blake2013tropical}. The trend continued with Hurricane Ida in 2021 causing power losses for 1.2 million customers across eight states, and the Texas cold snap affecting over 4.5 million customers. Beyond North America, Storm Ciarán in November 2023 left over a million people in France without electricity, with winds exceeding 200 kph causing widespread infrastructure damage \cite{cnn_storm_ciaran_2023}. These incidents, alongside the 2019 UK blackout that affected over one million people due to system inertia challenges with renewable integration, demonstrate that resilience failures result in substantial economic setbacks, threaten public safety, and disrupt essential services \cite{BIALEK2020111821}.

Despite growing recognition of resilience importance, current market mechanisms fail to provide adequate economic incentives for resilience resource investment. Existing capacity and reserve markets are designed primarily for normal operational conditions, with demand determined by load forecasts and conventional contingency planning rather than extreme event scenarios. Even when scarcity pricing occurs due to reserve insufficiency, the resulting prices reflect only normal operational value, failing to capture the true economic worth of resilience resources during extreme events. While localized pricing for dynamic microgrids has shown that emergency power can be delivered economically via intentional islanding, these ad-hoc models fail to create system-wide investment incentives \cite{SHANG2017321}. Price caps in most electricity markets further weaken these signals.



The economic burden of resilience enhancement is substantial. Recent studies demonstrate a quantifiable consumer willingness-to-pay (WTP) for fortification—estimated, for example, at nearly \$15/month in Oklahoma—yet this value varies significantly based on factors like residential location and political ideology \cite{Baik2020,LAMBERT2024107345}. This disparity, combined with the absence of transparent price signals that explicitly capture resilience value, creates inefficiencies in resource allocation and insufficient investment in resilience capabilities. Current market practices, such as post-2021 freeze modifications of Electric Reliability Council of Texas (ERCOT) \cite{ercot2021roadmap} and PJM's capacity market improvements, attempt to enhance resilience indirectly by elevating adequacy of existing market products \cite{9917337}, but these approaches fail to address the fundamental mismatch between resilience requirements and market incentives.

This paper addresses the urgent need for a comprehensive mathematical framework that enables the commercialization of energy resilience features in electrical grid planning and operation. The proposed framework aims to bridge the gap between resilience requirements and market mechanisms by developing transparent price signals that optimize resilience resource allocation, provide rational cost distribution among customers based on their resilience benefits, and incentivize investment in resilience enhancement. By establishing a value-driven market mechanism for resilience provision, this work seeks to transform how power systems prepare for and respond to extreme events, ensuring a more secure and stable power supply in an era of increasing environmental uncertainty and renewable energy integration.

The framework's development is particularly timely given the convergence of climate change impacts, aging infrastructure, cybersecurity threats, and the energy transition's operational challenges. As power systems become increasingly complex and interdependent, the ability to quantify, price, and trade resilience services becomes essential for maintaining grid stability and ensuring economic efficiency in resilience investments. This research contributes to the emerging field of resilience economics by providing practical tools for market operators, regulators, and system planners to implement resilience-aware market designs that balance cost-effectiveness with security of supply.

The provided methodology implements a novel Proximal Policy Optimization (PPO)-based reinforcement learning framework for dynamic grid switching optimization, with a focus on resilience and cost efficiency under varying weather conditions. Below are key the contributions:
\begin{enumerate}
    \item \textbf{Hierarchical Agent Structure}: The framework employs a dual-agent approach with a strategic PPO agent selecting high-level grid configurations and a tactical PPO agent optimizing individual switch states, enabling fine-grained control and adaptability to dynamic grid conditions.
    \item \textbf{Weather-Aware Decision Making}: The methodology simulates dynamic weather transitions (normal to calamity) with probabilistic calamity events, and the agents incorporate weather conditions into their state space, prioritizing resilience during calamities and cost efficiency during normal conditions.
    \item \textbf{Enhanced Reward Mechanism}: The reward function is designed to heavily incentivize high-resilience configurations during calamities (with bonuses up to 200x) while balancing cost efficiency in normal conditions, ensuring adaptive and context-sensitive decision-making.
    \item \textbf{Comprehensive Cost Modeling}: The Enhanced Cost Calculator integrates detailed economic analysis, including capital, operational, and failure costs, alongside revenue potential and risk reduction benefits, providing a realistic commercialization perspective for grid operations.
\end{enumerate}

The remainder of the paper is organized as follows. Section II shows the background work and literature review. Section III describes a comprehensive
framework for commercialization of resilience feature. A detailed case study and results is presented in Section IV. Section V concludes the paper by summarizing key findings, discussing broader implications, and suggesting potential directions for future research.

\section{Power grid resilience}

\subsection{Background work}

The concept of resilience in electrical distribution systems (EDS) encompasses a broad spectrum of capabilities, including risk anticipation, adaptability, reliability, robustness, and recovery. It is especially vital in addressing low-probability, high-impact events that may severely disrupt power delivery and infrastructure integrity.

The critical distinction between reliability and resilience further complicates the commercialization challenge. While reliability focuses on continuous power delivery under normal conditions, resilience encompasses the system's ability to prepare for, respond to, adapt to, and recover from adverse events. This broader scope requires new market mechanisms that can anticipate extreme events, optimize resilience resource allocation, and provide transparent pricing signals that reflect the true value of resilience services.

A widely accepted representation of this concept is the resilience trapezoid, which visualizes the system’s performance across different phases of a disruptive event—namely, the prevention, absorption, degradation, recovery, and adaptation stages \cite{Panteli201558,hussain2019microgrids}. This geometric interpretation offers a dynamic view of resilience by measuring both the depth and duration of performance degradation and the efficiency of recovery.

\begin{figure}[H]
    \centering
    \includegraphics[width=0.9\linewidth]{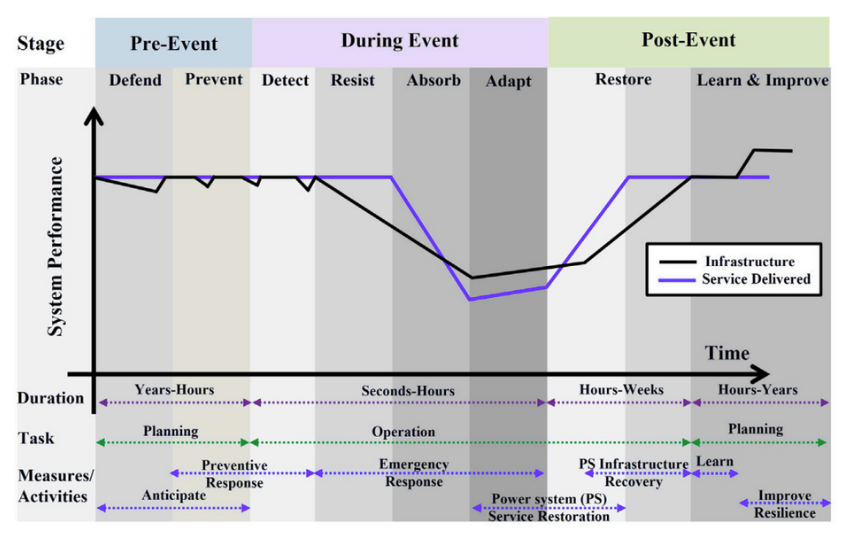}
    \caption{General Curve of System response to extreme event}
    \label{trapezoid_curve}
\end{figure}

The objective of a resilient system is not only to return to its pre-event operational state, but also to improve system attributes so that it becomes less vulnerable to future disruptions \cite{9913670}. This perspective shifts resilience from a purely reactive property to a proactive and adaptive capability.

Moreover, the resilience trapezoid as shown in Fig. \ref{trapezoid_curve} has been expanded to consider two interrelated curves:
\begin{itemize}
    \item One capturing infrastructure robustness, including network topology and component tolerance.
    \item The other measuring service continuity, particularly in maintaining supply to critical loads.
\end{itemize}

In 2022, the IEEE Task Force provided a formal definition of power system resilience as the ability to limit the extent, impact, and duration of system degradation to ensure the continued delivery of critical services following extraordinary events such as natural disasters, equipment failures, or cyber-attacks \cite{9913670}. This definition emphasizes not only enduring disruptions but also the system's capacity to anticipate, absorb, recover, adapt, and learn from adverse conditions.

This study adopts the IEEE definition as the foundation for evaluating the resilience of power distribution networks, focusing particularly on the ability to sustain supply to critical loads (CLs) and recover efficiently after disruption.

\subsection{Literature review}

Traditionally, the evaluation of electrical distribution system (EDS) performance has been grounded in established reliability metrics, including the System Average Interruption Duration Index (SAIDI), System Average Interruption Frequency Index (SAIFI), and Momentary Average Interruption Frequency Index (MAIFI) \cite{DWIVEDI2023156}. These conventional metrics are designed to assess the system's overall effectiveness in delivering power to all connected loads under normal operating conditions. However, these traditional reliability indicators prove inadequate for evaluating system performance during extreme events, where prioritizing critical loads over non-critical ones becomes essential \cite{PANWAR2019314, 9913670}.

The concept of "resilience" has emerged as a complementary assessment framework that specifically addresses the system's capacity to maintain uninterrupted power supply to critical loads during severe and unfavorable events. This paradigm shift recognizes that during extreme adversity, the focus must transition from serving all loads equally to ensuring the survival of critical infrastructure and services.

The transition toward renewable energy integration has introduced additional complexity to power system operations. The 2019 UK power outage, triggered by continuous tripping of gas and wind generators amid insufficient system inertia, exemplifies how the displacement of traditional thermal generation challenges secure and reliable operation \cite{BIALEK2020111821}. This has prompted market operators like Midcontinent Independent System Operator (MISO) and California Independent System Operator (CAISO) to introduce flexible ramping products in ancillary service markets, while inertia and primary frequency response markets are gaining widespread interest to address frequency stability issues in high-penetration renewable systems \cite{7458886,CORNELIUS20182291}.

The increasing occurrence of extreme events such as natural disasters and cyber-attacks has underscored the need for robust mechanisms that enhance power system resilience. Traditional electricity markets focus primarily on optimizing operations under normal conditions, offering limited scope for explicitly valuing and incentivizing resilience resources. Addressing this gap, Xiao et al. (2025) introduced a dedicated market mechanism for power system resilience provision \cite{XIAO2025125335}. Their model proposes a pre-event market where customers bid for the right to retain load during potential disruptions, reflecting their willingness to pay for prioritized recovery. The market is cleared based on robust optimization models that account for various failure scenarios, and the resulting resilience prices act as transparent signals for both customers and resilience providers. However, several limitations remain in their approach. The reliance on predefined failure scenarios introduces challenges in scalability and may not capture the full spectrum of possible disruptions. The computational complexity increases with the number of scenarios considered, making real-world implementation demanding. Importantly, Xiao et al. (2025) focus on static market-clearing mechanisms using robust optimization but do not explore adaptive or learning-based strategies that could respond to evolving grid conditions and stakeholder behaviors. This presents an opportunity for integrating advanced decision-making tools such as reinforcement learning (RL).

In this context, the present study extends the existing literature by proposing a Multi-Agent Proximal Policy Optimization (MAPPO) based framework for resilience-driven grid operation and planning. Unlike traditional market models, the RL-based approach dynamically learns optimal operational strategies and resource allocations, accounting for changing grid states, customer behaviors, and resilience priorities. This allows for continuous adaptation and potentially reduces computational overhead by avoiding explicit scenario enumeration. The integration of learning agents in resilience commercialization offers a novel pathway for both operational optimization and economic market design.

\subsection{Resilience Metric}

\begin{figure}[H]
    \centering
    \includegraphics[width=0.9\linewidth]{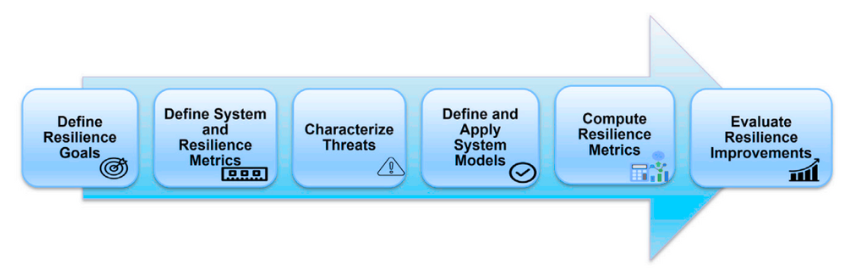}
    \caption{Sequential Stages of RAP}
    \label{rap}
\end{figure}

While numerous metrics have been proposed to quantify resilience, most do not align with the Resilience Analysis Process (RAP) proposed by Sandia National Laboratories \cite{watson2014conceptual}. The RAP outlines a six-step methodology for resilience assessment, guiding decision-making in both operational and planning domains as shown in Fig.\ref{rap}. A comprehensive resilience metric should be Quantitative and qualitative, scalable and comparable, risk-aware and uncertainty-resilient, and applicable across both short-term operations and long-term investment planning.

Reference \cite{DWIVEDI2025125001}, \cite{BABU2025126141} work integrates the IEEE definition and RAP framework to ensure that the proposed resilience metric and control strategy are both standards-compliant and actionable for real-world grid management.

\subsubsection{Parameters for electrical service requirements}

\paragraph{Weighted path variability}

The resilience evaluation of electrical distribution systems focuses on the availability of multiple supply paths to critical loads from grid and distributed energy sources. Network switching configurations create different supply scenarios categorized into three distinct types based on connectivity and redundancy characteristics.

\begin{itemize}
    \item Isolated paths ($N_{IP}$) occur when individual critical loads are supplied by single sources without alternative routing options.
    \item Isolated combinations ($N_{IC}$) involve multiple sources supplying all critical loads through separate, non-interconnected pathways.
    \item Connected combinations ($N_{CC}$) represent the most resilient configuration where multiple sources supply all critical loads through interconnected networks, providing maximum operational flexibility.
\end{itemize}
  
\noindent The weighting scheme reflects resilience preference hierarchy: 
\begin{itemize}
    \item Isolated paths receive minimal weight (0.1) due to limited backup capabilities.
    \item Isolated combinations are assigned moderate weight (0.4) for improved redundancy despite lack of interconnection.
    \item Connected combinations receive highest weight (0.5) for superior adaptability and recovery potential.
\end{itemize}

The weighted path variability is computed as:

\begin{equation}
PV_{CL} = 0.1 \cdot N_{IP} + 0.4 \cdot N_{IC} + 0.5 \cdot N_{CC}
\end{equation}

This metric quantifies supply path diversity and serves as a technical foundation for establishing tiered resilience services and pricing mechanisms within the commercialization framework.

\paragraph{Ratio of critical loads served}


The ratio of critical loads served ($N_{CLS}$) quantifies the fraction of critical load demand that can be successfully supplied by the system under various operational scenarios.


This metric is mathematically formulated as:

\begin{equation}
N_{CLS} = \frac{P_{CL}}{P_{TCL}}
\end{equation}

where $P_{CL}$ represents the actual power delivered to critical loads and $P_{TCL}$ denotes the aggregate power requirement of all critical loads in the system.

This ratio indicates the system's effectiveness in prioritizing essential load supply during both normal operations and contingency situations. The metric serves as a fundamental performance indicator for resilience assessment and forms a critical component in establishing differentiated service guarantees and corresponding pricing mechanisms within the proposed commercialization framework.

\paragraph{Average rating of service}

Average Rating of Service (ARoS) measures the system’s capacity to serve critical loads plus additional demand during extreme events. It reflects both resilience and operational flexibility, supporting tiered pricing.
The Rating of Service (RoS) is calculated for each supply configuration as:
\begin{equation}
RoS = \sum_{i=1}^{N_{PC}} \left[\frac{R_{Source} - R_{CL}}{R_{Source}}\right]_i
\end{equation}
where $R_{Source}$ is the available source rating (kVA), $R_{CL}$ is the critical load rating (kVA), and $N_{PC}$ is the number of possible supply paths.
The Average Rating of Service is then computed as:
\begin{equation}
ARoS = \frac{N_{CL}}{N_{TCL}} \times \frac{RoS}{N_{PC}}
\end{equation}
where $N_{CL}$ is the number of critical loads supplied and $N_{TCL}$ is the total critical loads.
A higher value resulting from this evaluation signifies an additional capacity to fulfill extra load demand. In other words, the system not only meets its essential operational requirements but also has the surplus capacity to accommodate additional loads, ensuring flexibility in its performance.

\subsubsection{Parameters for topological characteristics}

\paragraph{Percolation threshold for topological network}

The percolation threshold represents a critical probability point in percolation theory that marks the transition between isolated and connected phases in network systems, providing crucial insights into network resilience assessment. In percolation theory, nodes or edges are randomly assigned with probability $p$, and the percolation threshold $p_m$ identifies the critical point where an infinite spanning cluster emerges.

The percolation strength is calculated as:
\begin{equation}
P_{\infty}(p) = \frac{1}{NT} \sum_{i=1}^{T} S_i(p)
\end{equation}
where $N$ is the number of nodes in the network, $T$ is the total number of independent realizations of the Monte Carlo simulation, and $S_i(p)$ is the size of the largest cluster in the network during the $i$-th realization when the bond occupation probability is $p$.
The susceptibility, measuring fluctuations in cluster size, is defined as:
\begin{equation}
\chi(p) = \frac{\left[\frac{1}{N^2T} \sum_{i=1}^{T} S_i(p)^2\right] - [P_{\infty}(p)]^2}{P_{\infty}(p)}
\end{equation}
The percolation threshold is determined as:
\begin{equation}
p_m = \arg\max_p \chi(p)
\end{equation}
where susceptibility reaches its maximum, indicating the highest level of fluctuation in the largest cluster size. A higher percolation threshold indicates enhanced network resilience, as the system can withstand multiple failures without significant connectivity loss, while a lower threshold signifies network fragility prone to disconnection with minimal disruptions.

\paragraph{Information centrality for system’s nodes}

Information centrality identifies critical nodes essential for maintaining system functionality and ensuring reliable power supply to critical loads. For a network graph $G$ with $N$ nodes, the network efficiency $E_n[G]$ is calculated as:

\begin{equation}
E_n[G] = \frac{1}{N \cdot (N-1)} \sum_{i=1}^{N} \sum_{j \neq i} \frac{1}{d_{ij}}
\end{equation}

where $d_{ij}$ represents the shortest path distance between nodes $i$ and $j$.

The information centrality $C_m$ for node $m$ is determined by:

\begin{equation}
C_m = \frac{E_n[G] - E_n[G_0]}{E_n[G]}
\end{equation}

where $E_n[G]$ is the original network efficiency and $E_n[G_0]$ is the network efficiency after removing node $m$.

This metric quantifies the impact of individual node failures on network performance, enabling identification of critical infrastructure components whose protection and redundancy are essential for maintaining resilience services in the commercialization framework.

\subsubsection{Resilience Computation}

A comprehensive evaluation of resilience in electrical distribution systems (EDS) requires an approach that integrates both functional performance and topological characteristics. Prior studies \cite{DWIVEDI2024110537} have proposed multi-dimensional metrics that consider the system’s ability to deliver critical loads, maintain network connectivity, and withstand cascading failures under adverse conditions.

The resilience metric \( R \) is computed as a weighted sum of normalized indicators:

\begin{equation}
R = \mathbf{R} \cdot \mathbf{W}
\end{equation}

where \( \mathbf{R} = [PV_{CL}, N_{CLS}, A_{RoS}, p_m, N_{HC}] \) represents the vector of normalized resilience parameters, and \( \mathbf{W} = [w_{PV_{CL}}, w_{N_{CLS}}, w_{A_{RoS}}, w_{p_m}, w_{N_{HC}}] \) contains their respective weights.

\begin{itemize}
    \item \( PV_{CL} \): Path variability to critical loads
    \item \( N_{CLS} \): Number of critical loads served
    \item \( A_{RoS} \): Average rating of service
    \item \( p_m \): Percolation threshold (network vulnerability)
    \item \( N_{HC} \): Number of high-information-centrality nodes
\end{itemize}

Each parameter is normalized using min–max scaling to the range [0, 1], and the weights are derived using the \textbf{Analytic Hierarchy Process (AHP)}. The AHP involves pairwise comparisons of criteria to determine their relative importance, validated through consistency index and eigenvector computations.

To compare multiple switch configurations, a composite resilience score \( R_C \) is computed as:

\begin{equation}
R_C = R_{\text{max}} + (1 - R_{\text{max}}) \sum_{a=1}^{n-1} w_a R_a
\end{equation}

where \( R_{\text{max}} \) is the resilience of the best-performing configuration, \( R_a \) is the resilience score of the \( a \)-th configuration, and \( w_a \) is its assigned weight. Equal weights are assumed for simplicity in many scenarios.

The resilience evaluation framework incorporates the following operational constraints to ensure power system feasibility:

\begin{itemize}
    \item {Voltage constraints}:
    \begin{equation}
    |V_{\min}| \leq |V_n| \leq |V_{\max}|
    \end{equation}
    
    \item {Branch current limits}:
    \begin{align}
    |I_{f,n}| &\leq |I^{\max}_{f,n}| \\
    |I_{b,n}| &\leq |I^{\max}_{b,n}|
    \end{align}
    
    \item {Distributed Generation (DG) limits}:
    \begin{align}
    P^{\min}_{DG} \leq P_{DG} &\leq P^{\max}_{DG} \\
    Q^{\min}_{DG} \leq Q_{DG} &\leq Q^{\max}_{DG}
    \end{align}
\end{itemize}

Here, \( V_n \) is the node voltage, \( I_{f,n} \) and \( I_{b,n} \) are forward and backward branch currents, and \( P_{DG} \), \( Q_{DG} \) represent the real and reactive power of distributed generators. Subscripts \( \min \) and \( \max \) denote operational limits.


\noindent This established metric forms the foundational layer for our framework, where we integrate it with a \textbf{multi-agent deep reinforcement learning (DRL)} architecture based on Proximal Policy Optimization (PPO). Our proposed system extends resilience evaluation to a real-time, adaptive, and economically-aware decision-making model for distribution network operation and planning.

\begin{figure}[H]
    \centering
    \includegraphics[width=0.9\linewidth]{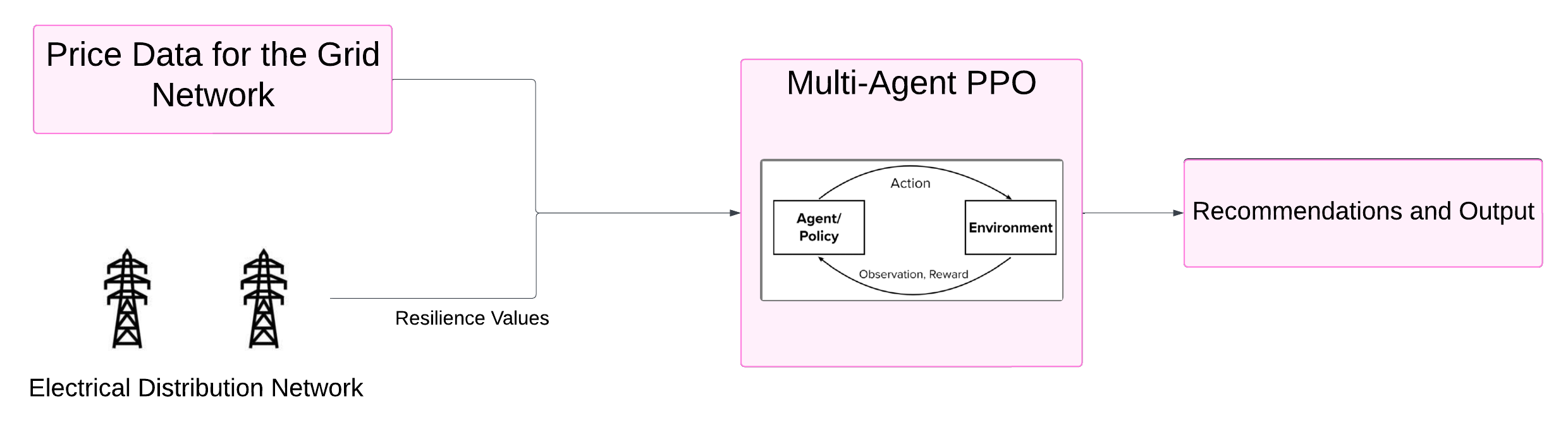}
    \caption{System Architecture of Agentic-AI based Recommendation System}
    \label{System_diagram}
\end{figure}

\subsection{Strategies to improve resilience in EDS}

Enhancing the resilience of electrical distribution systems (EDS) involves both planning and operational strategies that strengthen the system’s ability to withstand, absorb, and recover from various disruptions. Among the most effective methods is the integration of distributed energy resources (DERs), which improve reliability, reduce blackout impacts, and optimize local power supply \cite{10086607}.

As DER penetration grows within centralized grids, strategic allocation becomes essential to enhance voltage stability, minimize losses, and improve overall reliability \cite{MEHIGAN2018822,DWIVEDI2024100696}. Numerous studies have proposed optimization models for DER placement, focusing on minimizing energy losses, handling variability in generation patterns, and reducing investment and maintenance costs \cite{UNIYAL2021106909,Haimes2009,Naderi20121313}. Further enhancements include the use of diverse DER technologies and the development of microgrids to increase supply security \cite{Arefifar20131567}.

Energy storage systems, both fixed \cite{WANG2023108920,WANG2023128677} and mobile \cite{10078417,9035499}, also play a vital role. Recent advancements in AI-driven battery management have extended the operational life of lithium-ion batteries, enabling storage systems to support resilience through energy backup and load balancing \cite{MA2024512,HASIB2024100003}.

Another crucial strategy involves the deployment of automated tie-line switches, which improve system flexibility by allowing fast reconfiguration during faults. These switches facilitate isolation of damaged segments and power rerouting, thereby enabling quicker restoration. While past studies have focused on switch placement for reliability \cite{Heidari20151401,Celli19991167}, the work \cite{DWIVEDI2025125001} emphasizes resilience-centric planning, integrating both DERs and automated switching to improve the system’s capacity to anticipate, absorb, and recover from high-impact disruptions.


\section{Materials and Method}

\subsection{Considered System}

\begin{figure}[H]
    \centering
    \includegraphics[width=0.9\linewidth]{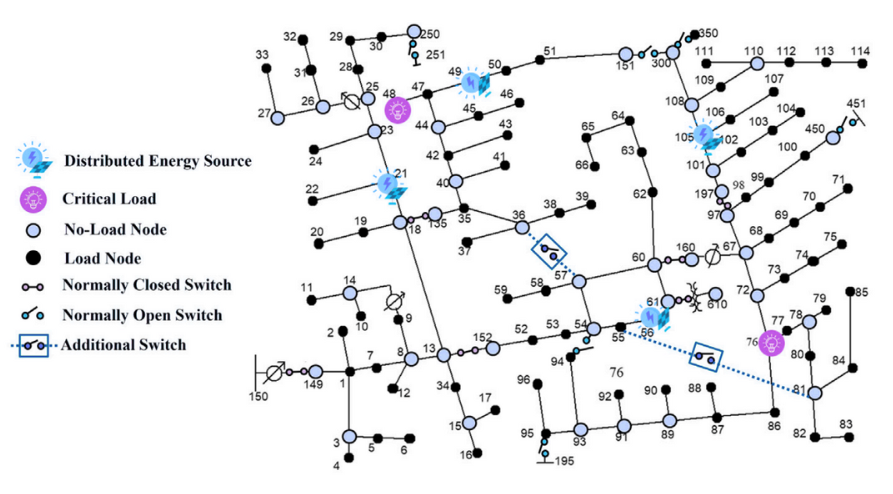}
    \caption{Line Diagram of IEEE 123 node test feeder, with integration of additional switches and DERs \cite{DWIVEDI2025125001}}
    \label{ieee123}
\end{figure}


To validate the proposed commercialization framework for energy resilience features, the IEEE 123-node test feeder is employed as the analysis platform, as illustrated in Fig.\ref{ieee123}. This distribution system operates at a nominal voltage of 4.16 kV and is supplied by a main substation located at node 150 with a rated capacity of 5000 kVA. The system comprises 85 load nodes with a total power demand of 3855.26 kVA distributed throughout the network.

The IEEE 123-node test feeder offers several characteristics that make it particularly suitable for resilience analysis and commercialization framework development. The system incorporates 12 switching devices, consisting of 6 normally closed switches and 6 normally open tie switches, which enable network reconfiguration capabilities during extreme events. This switching infrastructure provides multiple alternate supply paths to maintain power delivery to critical loads when primary feeders are compromised, making it an ideal testbed for evaluating resilience strategies and their associated economic value.

To enhance the system's resilience capabilities and reflect modern distribution network trends, four distributed energy resources (DERs) are strategically integrated into the network. These DERs are positioned at nodes 49, 21, 105, and 56, each rated at 350.35 kVA, providing a total distributed generation capacity of 1401.4 kVA. The strategic placement of these resources enables local power supply during grid disruptions and contributes to the overall system resilience by reducing dependency on the main substation during extreme events.

For the commercialization framework analysis, two critical load nodes are designated based on their high power demand and essential service requirements. Node 48 and node 76 are identified as critical loads with power demands of 258.20 kVA and 303.69 kVA, respectively. These nodes collectively represent 561.89 kVA, accounting for approximately 15\% of the total system load demand. The selection of these nodes as critical loads is justified by their substantial power requirements and their representation of essential facilities such as hospitals, data centers, emergency services, and other infrastructure that requires continuous power supply during extreme events.

The designation of these critical loads serves multiple purposes within the commercialization framework as follows:
\begin{itemize}
    \item It establishes a clear hierarchy of load importance that can be translated into differentiated pricing structures for resilience services.
    \item It provides a basis for evaluating the economic value of maintaining power supply to high-priority customers during extreme events.
    \item It enables the assessment of how resilience resources should be allocated and priced to ensure optimal service delivery to critical infrastructure while maintaining economic efficiency across the entire distribution network.
\end{itemize}

The modified IEEE 123-node system thus provides a comprehensive platform for testing the proposed commercialization framework, incorporating the key elements necessary for resilience analysis: network reconfiguration capabilities, distributed generation resources, differentiated load criticality, and realistic operational constraints. This system configuration allows for a detailed evaluation of how resilience services can be quantified, priced, and traded in a market-based framework while ensuring a reliable power supply to essential services during extreme events.

\subsection{Hierarchical PPO for Grid Switching}

The grid switching problem is formulated as a hierarchical reinforcement learning task with two levels of control:

\begin{enumerate}
    \item \textbf{Strategic Level:} The high-level agent selects a grid configuration \( c \in \{0, 1, 2, 3, 4, 5\} \), representing predefined operational topologies or DER deployments, based on environmental context.

    \item \textbf{Tactical Level:} Given the selected configuration, a low-level agent optimizes the individual switch states \( \mathbf{s} = [s_1, s_2, \ldots, s_{10}] \), where \( s_i \in \{0, 1\} \), to maximize resilience and ensure all operational constraints are satisfied.
\end{enumerate}

The environment state at time step \( t \) is represented by the vector:

\[
\mathbf{x}_t = [w_t, s_1^t, \ldots, s_{10}^t, b_t, p_t, \mathbf{r}_t, \mathbf{f}_t] \in \mathbb{R}^{20}
\]

where \( w_t \) denotes the current weather condition, \( b_t \) is the normalized budget, \( p_t \) indicates the episode progress, \( \mathbf{r}_t \) represents resilience metrics, and \( \mathbf{f}_t \) includes contextual features such as cost efficiency and fault duration.

This hierarchical structure enables modular and interpretable control: the strategic agent determines the system’s overall operational posture, while the tactical agent performs fine-grained optimization of switch states within the chosen configuration.

\begin{figure}[h]
    \centering
    \includegraphics[width=0.9\linewidth]{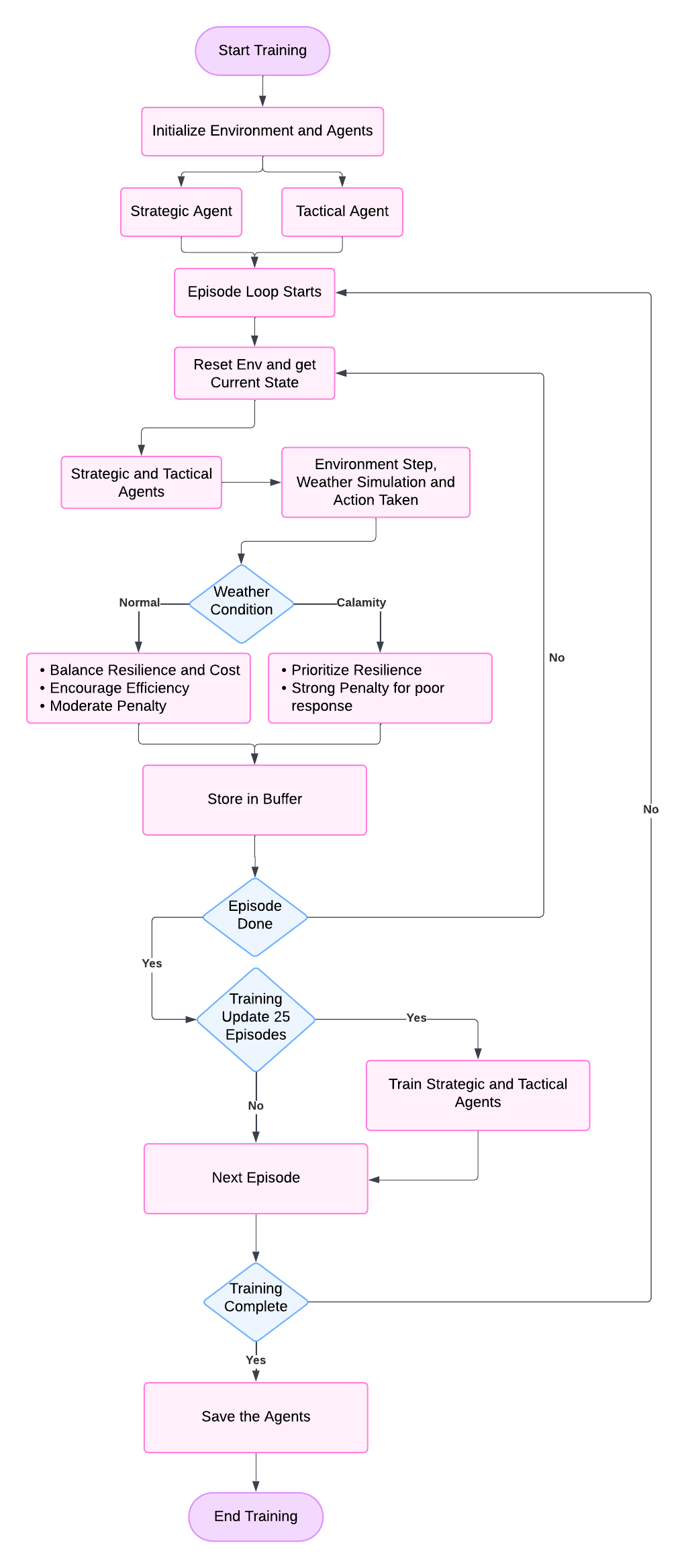}
    \caption{Algorithmic Workflow}
    \label{workflow}
\end{figure}

Fig.\ref{workflow} shows the training workflow starting with environment and agent initialization, followed by an episode loop. It involves resetting the environment, simulating weather and actions, and adjusting strategies based on normal or calamity conditions. Data is stored in a buffer, and training updates occur every 25 episodes. The process repeats until training is complete, saving the agents at the end.












\subsection{Agentic-AI}

\subsubsection{Strategic Agent Policy Design using PPO}

The strategic agent is responsible for selecting the optimal switch configuration from a set of predefined actions under varying grid conditions. It is trained using the \textit{Proximal Policy Optimization (PPO)} algorithm, a stable and effective policy-gradient method for continuous learning in complex environments.

The agent learns a policy $\pi_{\theta_s}(c \mid s)$, where $c$ denotes a switch configuration and $s$ represents the observed system state. The action probabilities are defined as:

\begin{equation}
\pi_{\theta_s}(c \mid s) = \text{Softmax}(h_s + b_{\text{emergency}})
\end{equation}

Here, $h_s$ is the base output (logits) from a feedforward actor network, and $b_{\text{emergency}}$ is an adaptive bias term used during adverse weather conditions. The emergency bias is computed as:

\begin{equation}
b_{\text{emergency}} = w_t \cdot \boldsymbol{\alpha}, \quad \boldsymbol{\alpha} = [0.0, 0.5, 1.0, 2.0, 4.0, 5.0]
\end{equation}

where $w_t \in \{0, 1\}$ indicates the weather condition (0 for normal, 1 for calamity), and $\boldsymbol{\alpha}$ is a predefined weighting vector to prioritize more resilient configurations during emergencies.

An action $c_t$ is sampled from the resulting categorical distribution as follows:

\begin{equation}
c_t \sim \text{Categorical}(\pi_{\theta_s}(\cdot \mid s_t))
\end{equation}

The actor is trained using the clipped surrogate PPO loss with an entropy regularization term to maintain exploration. A separate critic network $V(s)$ estimates the value of each state, and the Generalized Advantage Estimation (GAE) method is used to compute temporal credit assignment.

The integration of weather-aware bias and PPO training enables the strategic agent to dynamically adapt its configuration policy in response to both environmental and operational conditions, improving resilience decision-making during extreme events.










\subsubsection{Tactical Agent Policy for Switch-Level Control}

The tactical agent operates at a finer control granularity, taking as input the joint state and strategic decision to optimize individual switch states and grid preferences. The input vector to the tactical policy is defined as:

\begin{equation}
x^{\text{tactical}}_t = [x_t, c_t] \in \mathbb{R}^{21}
\end{equation}

where $x_t$ represents the environment state and $c_t$ is the action selected by the strategic agent. The tactical actor network produces two independent output heads: one for binary switch control decisions, and another for grid selection.

\vspace{0.5em}
\noindent\textbf{Switch Control Head:} The probability of turning ON each switch $s_i$ is defined as:

\begin{equation}
\pi^{\text{switch}}_{\theta_t}(s_i \mid s) = \sigma(h^{\text{switch}}_i + b^{\text{emergency}}_i), \quad b^{\text{emergency}}_i = w_t \cdot 2.0
\end{equation}

Here, $\sigma(\cdot)$ denotes the sigmoid function, $w_t$ is the weather indicator (1 during calamity), and $h^{\text{switch}}_i$ is the $i^{\text{th}}$ switch logit. This bias ensures higher switch closure probabilities during emergencies, improving fault-tolerant connectivity.

\vspace{0.5em}
\noindent\textbf{Grid Selection Head:} A separate grid decision is made as:

\begin{equation}
\pi^{\text{grid}}_{\theta_t}(g \mid s) = \sigma(h^{\text{grid}})
\end{equation}

For binary selection (i.e., one grid option), the sigmoid output serves as the probability of choosing that grid configuration. For multiple options, a softmax function is used.

\vspace{0.5em}
\noindent\textbf{Joint Action Probability:} The overall probability of the selected tactical action is the product of switch-wise probabilities and the grid decision:

\begin{equation}
\pi_{\theta_t}(a_t \mid x^{\text{tactical}}_t) = \left( \prod_{i=1}^{10} \pi^{\text{switch}}_{\theta_t}(s_i \mid x^{\text{tactical}}_t) \right) \cdot \pi^{\text{grid}}_{\theta_t}(g_t \mid x^{\text{tactical}}_t)
\end{equation}

This formulation enables simultaneous optimization of physical topology (via switches) and logical resource use (via grid selection).

\vspace{0.5em}
\noindent\textbf{Learning Procedure:} The tactical agent is trained using PPO with clipped surrogate loss, entropy regularization, and a value function loss. The policy captures spatiotemporal dependencies while adapting to emergency conditions. The emergency-aware bias significantly improves adaptive behavior under stress, enabling granular response decisions that complement the strategic agent.

\begin{algorithm}
\label{training}
\caption{Training of Dynamic Switching Agent}
\begin{algorithmic}[1]
    \State \textbf{Input:} $\mathcal{D}$ (Grid data), $\tau$ (Checkpoint interval), $\sigma$ (Save best only flag), $\kappa$ (Number of checkpoints to keep), $\rho$ (Resume checkpoint path)
    \State \textbf{Output:} $\mathcal{A}_s$ (Trained strategic agent), $\mathcal{A}_t$ (Trained tactical agent), $\mathcal{M}$ (Training metrics)
    
    \State Initialize random seeds: $\mathcal{R}_s$
    \State Create environment: $\mathcal{E} \gets \text{Env}(\mathcal{D}, l=100)$
    \State Initialize strategic agent: $\mathcal{A}_s \gets \text{Agent}_s(d_s=20, d_a=6)$
    \State Initialize tactical agent: $\mathcal{A}_t \gets \text{Agent}_t(d_s=21, d_{sw}=10, d_g=1)$
    \State Initialize experience buffers: $\mathcal{B}_s$, $\mathcal{B}_t$
    \State Set parameters: $\epsilon \gets 1200$, $\phi \gets 25$
    \State Load checkpoint: $\text{Load}(\mathcal{A}_s, \mathcal{A}_t, \rho)$ if $\rho \neq \emptyset$
    
    \For{\textbf{each} $\epsilon_i \in [\epsilon_{\text{start}}, \epsilon]$}
        \State Reset environment: $\mathcal{S}_0 \gets \mathcal{E}.\text{Reset}()$
        \State Initialize tracking: $\mathcal{T} \gets \emptyset$
        
        \While{\textbf{not} $\delta$}
            \State Strategic action: $\mathcal{A}_s \to (\alpha_s, p_s, v_s)$
            \State Enhance state: $\mathcal{S}_t \gets \mathcal{S} \oplus \alpha_s$
            \State Tactical action: $\mathcal{A}_t \to (\alpha_{sw}, \alpha_g)$
            \State Step environment: $(\mathcal{S}', r, \delta, \mathcal{I}) \gets \mathcal{E}.\text{Step}(\alpha_s, \alpha_{sw}, \alpha_g)$
            \State Store experiences: $\mathcal{B}_s \gets \mathcal{B}_s \cup \{(\mathcal{S}, \alpha_s, p_s, r, v_s)\}$, $\mathcal{B}_t \gets \mathcal{B}_t \cup \{(\mathcal{S}_t, \alpha_{sw}, \alpha_g, r)\}$
            \State Update state: $\mathcal{S} \gets \mathcal{S}'$, $\mathcal{T} \gets \mathcal{T} \cup r$
        \EndWhile
        
        \State Record metrics: $\mathcal{M} \gets \mathcal{M} \cup \mathcal{T}$
        \If{$(\epsilon_i + 1) \mod \phi = 0 \land \mathcal{B}_s, \mathcal{B}_t \neq \emptyset$}
            \State Train strategic: $\mathcal{A}_s.\text{Train}(\mathcal{B}_s)$
            \State Train tactical: $\mathcal{A}_t.\text{Train}(\mathcal{B}_t)$
            \State Clear buffers: $\mathcal{B}_s, \mathcal{B}_t \gets \emptyset$
        \EndIf
        
        \If{$(\epsilon_i + 1) \mod \tau = 0$}
            \State Save checkpoint: $\text{Save}(\mathcal{A}_s, \mathcal{A}_t, \epsilon_i, \mathcal{M}, \sigma)$
            \State Cleanup checkpoints: $\text{Cleanup}(\kappa)$
        \EndIf
    \EndFor
    
    \State Save final checkpoint: $\text{Save}(\mathcal{A}_s, \mathcal{A}_t, \epsilon, \mathcal{M}, \text{False})$
    \State Return $\mathcal{A}_s$, $\mathcal{A}_t$, $\mathcal{M}$
\end{algorithmic}
\end{algorithm}

\subsection{Resilience-Oriented Reward Shaping and Environment Design}

The environment simulates dynamic grid conditions under both normal and extreme weather events (calamities), enabling multi-agent training using a reward structure sensitive to resilience, cost, and adaptation. Two distinct reward functions are defined:

\textbf{a) Normal Conditions:}
\begin{equation}
R^{\text{normal}}_t = \alpha_1 \cdot \rho_t - \alpha_2 \cdot \text{cost}(c_t, s_t) + \beta_{\text{efficiency}}
\end{equation}

\textbf{b) Calamity Conditions:}
\begin{equation}
R^{\text{calamity}}_t = \alpha_3 \cdot \rho_t - \alpha_4 \cdot \text{cost}(c_t, s_t) + \beta_{\text{resilience}} + \beta_{\text{adaptation}}
\end{equation}

Where $\rho_t$ represents the total resilience at time $t$, $\text{cost}(c_t, s_t)$ is the configuration and switching cost, and $\beta$ terms are additional bonuses. The resilience bonus $\beta_{\text{resilience}}$ is tiered as follows:

\[
\beta_{\text{resilience}} =
\begin{cases}
+150, & \text{if } \rho_t > 0.8 \\
+100, & \text{if } 0.7 < \rho_t \leq 0.8 \\
+50, & \text{if } 0.6 < \rho_t \leq 0.7 \\
-100, & \text{if } \rho_t \leq 0.6
\end{cases}
\]





This accounts for tactical switch activation, boosting resilience based on the number of closed switches.

\textbf{Reward Normalization:} Rewards are normalized to stabilize learning:

\begin{equation}
\tilde{R}_t = \frac{R_t - \mu_R}{\sigma_R}
\end{equation}

\textbf{Training Objectives:} Both strategic and tactical agents are optimized using PPO with clipped loss, value error, and entropy regularization:

\begin{equation}
\mathcal{L}_{\text{strategic}} = \mathcal{L}_{\text{CLIP}}(\theta_s) - c_1 \mathcal{L}_{\text{VF}}(\phi_s) + c_2 \mathcal{L}_{\text{ENT}}(\theta_s)
\end{equation}
\begin{equation}
\mathcal{L}_{\text{tactical}} = \mathcal{L}_{\text{CLIP}}(\theta_t) - c_1 \mathcal{L}_{\text{VF}}(\phi_t) + c_2 \mathcal{L}_{\text{ENT}}(\theta_t)
\end{equation}

The tactical agent includes modified entropy over both switch and grid actions:

\begin{equation}
\mathcal{L}_{\text{ENT}}(\theta_t) = -\mathbb{E}_t \left[ \sum_{i=1}^{10} H(s_i) + H(g) \right]
\end{equation}
\begin{equation}
H(s_i) = -p_i \log p_i - (1 - p_i) \log(1 - p_i)
\end{equation}

\textbf{Environment Dynamics:} The grid simulation includes weather transitions, cost constraints, and state evolution, where switch configurations and DER deployments are evaluated for resilience and cost. Each agent learns to make budget-constrained, resilience-optimized decisions in both stable and disrupted conditions.

\begin{table}[H]
\centering
\caption{Training Setup and PPO Hyperparameters}
\small 
\begin{tabular}{|p{3.5cm}|p{4.5cm}|}
\hline
\textbf{Training Design} & \textbf{Details} \\
\hline
Experience Collection & Parallel from both agents (strategic + tactical) \\
Reward Split & 60\% Strategic agent, 40\% Tactical agent \\
PPO Updates & Applied independently to each agent \\
Gradient Clipping & Max norm = 0.5 \\
KL Early Stopping & Triggered on divergence $>$ 1.5 $\times$ target KL \\
\hline
\textbf{Hyperparameter} & \textbf{Value} \\
\hline
Learning Rate (Strategic) & $3 \times 10^{-4}$ \\
Learning Rate (Tactical) & $3 \times 10^{-5}$ \\
Discount Factor ($\gamma$) & 0.99 \\
GAE Lambda ($\lambda$) & 0.95 \\
Clipping Parameter ($\epsilon$) & 0.2 \\
Value Loss Coefficient ($c_1$) & 0.5 \\
Entropy Coefficient ($c_2$) & 0.01 \\
Mini-Batch Size & 64 \\
PPO Epochs & 4 \\
Update Frequency & Every 25 episodes \\
\hline
\end{tabular}
\label{tab:ppo_hyperparams}
\end{table}

This hierarchical PPO architecture enables adaptive, weather-aware control over both strategic grid configuration and fine-grained switch settings, balancing resilience and cost under dynamic conditions. Algorithm 1 details the training of the dynamic switching agent with hierarchical strategic and tactical interactions for optimal grid configuration, while Algorithm 2 outlines the loading process to restore agent states and metadata.

\begin{algorithm}
\label{testing}
\caption{Loading Model Checkpoint}
\begin{algorithmic}[1]
    \State \textbf{Input:} $\mathcal{A}_s$ (Strategic agent), $\mathcal{A}_t$ (Tactical agent), $\pi$ (Checkpoint path)
    \State \textbf{Output:} $\zeta$ (Success flag), $\mu$ (Metadata)
    
    \State Attempt to load: $\mathcal{A}_s.\text{Actor}.\text{Load}(\pi_{\text{actor}})$
    \State Attempt to load: $\mathcal{A}_s.\text{Critic}.\text{Load}(\pi_{\text{critic}})$
    \State Attempt to load: $\mathcal{A}_t.\text{Actor}.\text{Load}(\pi_{\text{actor}})$
    \State Attempt to load: $\mathcal{A}_t.\text{Critic}.\text{Load}(\pi_{\text{critic}})$
    
    \If{all load attempts successful}
        \State Load metadata: $\mu \gets \text{Read}(\pi_{\text{meta}})$ \Comment{If $\pi_{\text{meta}}$ exists}
        \If{$\mu \neq \emptyset$}
            \State Extract: $\epsilon \gets \mu_{\text{episode}}$, $r_{\text{avg}} \gets \mu_{\text{reward}}$
            \State Restore statistics: $\mathcal{A}_s.\text{Stats} \gets \mu_s$
            \State Restore statistics: $\mathcal{A}_t.\text{Stats} \gets \mu_t$
            \State Restore normalization: $\mathcal{A}_s.\text{Norms} \gets \mu_{n_s}$
            \State Restore normalization: $\mathcal{A}_t.\text{Norms} \gets \mu_{n_t}$
        \EndIf
        \State Set success flag: $\zeta \gets \text{True}$
        \State Print success message: $\text{Success}(\epsilon, r_{\text{avg}})$
    \Else
        \State Set failure flag: $\zeta \gets \text{False}$
        \State Print error message: $\text{Error}(\text{load failure})$
    \EndIf
    
    \State \Return $\zeta$, $\mu$
\end{algorithmic}
\end{algorithm}

\subsection{Commercialization}

\begin{table*}[htbp]
\centering
\caption{Predefined Parameters for Commercial Analysis}
\begin{tabular}{l l r p{6cm}}
\toprule
\textbf{Parameter} & \textbf{Symbol} & \textbf{Value} & \textbf{Description} \\
\midrule
\multicolumn{4}{c}{\textbf{Capital Costs}} \\
Control system & $C_{\text{ctrl}}$ & 50000 & Cost of control system (\$) \\
DER unit & $C_{\text{DER}}$ & 3000 & Cost per DER unit (\$) \\
Switch installation & $C_{\text{sw}}$ & 1500 & Cost per switch installation (\$) \\
Communication system & $C_{\text{comm}}$ & 2500 & Cost per communication system (\$) \\
Protection equipment & $C_{\text{prot}}$ & 2000 & Cost per protection equipment (\$) \\
\midrule
\multicolumn{4}{c}{\textbf{Operational Costs}} \\
Maintenance per switch & $C_{\text{maint}}$ & 2 & Maintenance cost per switch (\$) \\
Fuel cost per DER & $C_{\text{fuel}}$ & 10 & Fuel cost per DER unit (\$) \\
Communication cost & $C_{\text{comm-op}}$ & 0.5 & Communication operational cost per switch (\$) \\
Operator cost & $C_{\text{operator}}$ & 30 & Operator cost per step (\$) \\
Monitoring cost & $C_{\text{monitor}}$ & 5 & Monitoring cost per step (\$) \\
\midrule
\multicolumn{4}{c}{\textbf{Failure Costs}} \\
Outage cost per customer & $C_{\text{outage}}$ & 5 & Cost per customer outage (\$) \\
Restoration cost per crew & $C_{\text{restore}}$ & 50 & Cost per restoration crew (\$) \\
Emergency generation & $C_{\text{emerg}}$ & 200 & Cost of emergency generation (\$) \\
Equipment damage & $C_{\text{damage}}$ & 1000 & Cost of equipment damage (\$) \\
Reputation cost & $C_{\text{rep}}$ & 5000 & Reputation cost per failure (\$) \\
\midrule
\multicolumn{4}{c}{\textbf{Configuration Multipliers}} \\
Capital multiplier (config 0--5) & $\mu_{\text{cap}}(c)$ & 1.0--3.5 & Multiplier for capital costs \\
Operational multiplier (config 0--5) & $\mu_{\text{op}}(c)$ & 1.0--2.0 & Multiplier for operational costs \\
Maintenance multiplier (config 0--5) & $\mu_{\text{maint}}(c)$ & 1.0--1.5 & Multiplier for maintenance costs \\
\midrule
\multicolumn{4}{c}{\textbf{Other Parameters}} \\
Discount rate & $\delta$ & 0.03 & Annual discount rate \\
Project lifetime & $L$ & 5 & Project duration (years) \\
Inflation rate & $\eta$ & 0.03 & Annual inflation rate \\
Number of customers & $N_c$ & 50 & Number of customers served \\
Value of outage & $V_o$ & 100 & Value of avoiding outage per customer (\$) \\
Subscription fee & $S_{\text{fee}}$ & 2000 & Monthly subscription fee per subscriber (\$) \\
Number of subscribers & $N_s$ & 1 & Number of subscribers \\
Performance contract & $P_{\text{contract}}$ & 1000 & Performance contract value per resilience unit (\$) \\
Incentive rate & $I_{\text{rate}}$ & 2000 & Incentive rate per resilience unit in calamity (\$) \\
Development cost & $C_{\text{dev}}$ & 30000 & Annual development cost (\$) \\
Initial cost & $C_{\text{init}}$ & 150000 & Initial investment cost (\$) \\
Penalty cost & $P_{\text{cost}}$ & 1000 & Penalty for low resilience in calamity (\$) \\
Baseline resilience & $r_{\text{base}}$ & 0.5 & Baseline resilience score \\
Configuration base cost & $C_{\text{config}}(c)$ & 0, 15, 28, 40, 50, 75 & Base cost for configurations 0--5 (\$) \\
Number of DER units & $n_{\text{DER}}(c)$ & 0, 1, 2, 3, 4, 4 & DER units for configurations 0--5 \\
Episode length & $T$ & 50 & Steps per episode \\
\bottomrule
\end{tabular}
\end{table*}

\subsubsection{Capital Cost}

\begin{equation}
\begin{split}
C_{\text{cap}}(c, s_t) = \Big( C_{\text{ctrl}} + n_{\text{DER}}(c) \cdot C_{\text{DER}} \\
+ s_t \cdot (C_{\text{sw}} + C_{\text{comm}} + C_{\text{prot}}) \Big) \cdot \frac{\mu_{\text{cap}}(c)}{L}
\end{split}
\end{equation}

It calculates the annualized capital cost, including control system, DER units, switches, communication, and protection equipment, scaled by a configuration-specific multiplier and amortized over the project lifetime.

\subsubsection{Operational Cost}
\begin{equation}
\begin{split}
C_{\text{op}}(c, s_t, w_t) = \Big( C_{\text{operator}} + C_{\text{monitor}} + C_{\text{maint}} \cdot s_t \\
+ n_{\text{DER}}(c) \cdot C_{\text{fuel}} + C_{\text{comm-op}} \cdot s_t \Big) \cdot \mu_{\text{op}}(c) \cdot \gamma(w_t)
\end{split}
\end{equation}

\textbf{where } $\gamma(w_t) = \begin{cases} 1.5 & \text{if } w_t = \text{Calamity} \\ 1.0 & \text{otherwise} \end{cases}$

It captures operational costs adjusted for configuration complexity and increases during calamities.

\subsubsection{Failure Cost}
\begin{equation}
\begin{split}
C_{\text{fail}}(r_t, w_t) = (1 - r_t) \cdot \Big( N_c \cdot C_{\text{outage}} + 2 \cdot C_{\text{restore}} \\
+ 10 \cdot C_{\text{emerg}} \Big) \cdot \phi(w_t)
\end{split}
\end{equation}

\textbf{where } $\phi(w_t) = \begin{cases} 3.0 & \text{if } w_t = \text{Calamity} \\ 1.0 & \text{otherwise} \end{cases}$

This models economic losses from outages, restoration, and emergency generation, amplified during calamities.

\subsubsection{Resilience Value}
\begin{equation}
V_{\text{res}}(r_t) = r_t \cdot V_o \cdot N_c
\end{equation}

It tells about benefit of maintaining resilience, scaled by reliability score and customer base.

\subsubsection{Revenue Potential}
\begin{equation}
\begin{split}
R_{\text{rev}}(r_t, w_t) = S_{\text{fee}} \cdot N_s \cdot \frac{T}{30} + P_{\text{contract}} \cdot r_t \\
+ I_{\text{rate}} \cdot r_t \cdot \mathbb{I}(w_t = \text{Calamity})
\end{split}
\end{equation}

\textbf{where } $\mathbb{I}(w_t = \text{Calamity}) = \begin{cases} 1 & \text{if calamity} \\ 0 & \text{otherwise} \end{cases}$

This shows the revenue from subscription, performance, and calamity incentives.

\subsubsection{Risk Reduction Benefit}
\begin{equation}
B_{\text{risk}}(r_t, w_t) = 
\begin{cases} 
P_{\text{cost}} \cdot (r_{\text{base}} - r_t), & \text{if } w_t = \text{Calamity}  \\ \text{ and } r_t < r_{\text{base}} \\
0, & \text{otherwise}
\end{cases}
\end{equation}

This penalizes resilience below baseline during calamity, rewarding reliability.

\subsubsection{Total Cost}
\begin{equation}
\begin{split}
C_{\text{total}} = \sum_{t=1}^T \Big( C_{\text{config}}(c_t) + C_{\text{cap}}(c_t, s_t) \\
+ C_{\text{op}}(c_t, s_t, w_t) + C_{\text{fail}}(r_t, w_t) \Big) + C_{\text{dev}} \cdot \frac{T}{8760}
\end{split}
\end{equation}

It is the sum of all cost components over time including config, capital, operations, failure, and development.

\subsubsection{Net Present Value (NPV)}
\begin{equation}
\text{NPV} = -C_{\text{init}} + \sum_{y=0}^{L-1} \frac{CF_y}{(1 + \delta)^y}
\end{equation}

\begin{equation}
\begin{split}
CF_y = \sum_{t \in \text{year } y} \Big( R_{\text{rev}}(r_t, w_t) \\
- ( C_{\text{config}}(c_t) + C_{\text{cap}}(c_t, s_t) + C_{\text{op}}(c_t, s_t, w_t) \\
+ C_{\text{fail}}(r_t, w_t) ) \Big)
\end{split}
\end{equation}

The NPV discounts future cash flows (revenue minus cost) over lifetime.

\subsubsection{Cost-Effectiveness Metrics}
\begin{itemize}
    \item {Benefit-Cost Ratio (BCR):}
    \begin{equation}
    \text{BCR} = \frac{R_{\text{rev,total}} + B_{\text{risk,total}}}{C_{\text{total}}}
    \end{equation}
    
    \item {Cost per Unit Benefit (CPUB):}
    \begin{equation}
    \text{CPUB} = \frac{C_{\text{total}}}{R_{\text{rev,total}} + B_{\text{risk,total}}}
    \end{equation}

    \item {Net Benefit (NB):}
    \begin{equation}
    \text{NB} = R_{\text{rev,total}} + B_{\text{risk,total}} - C_{\text{total}}
    \end{equation}
\end{itemize}

\textbf{where}:
\begin{align*}
R_{\text{rev,total}} &= \sum_{t=1}^T R_{\text{rev}}(r_t, w_t), \\
B_{\text{risk,total}} &= \sum_{t=1}^T B_{\text{risk}}(r_t, w_t)
\end{align*}

These metrics guide economic decisions by comparing costs against revenue and risk benefits.

\section{Results and Discussion}

Using the proposed methodology, the model was trained for recommendation and the following results were obtained for different coningencies.

\subsection{Flood}

Consider the distribution system operating under a flood scenario, where certain regions (e.g., the shaded zone) are affected by network isolation.

\begin{figure}[H]
    \centering
    \includegraphics[width=0.85\linewidth]{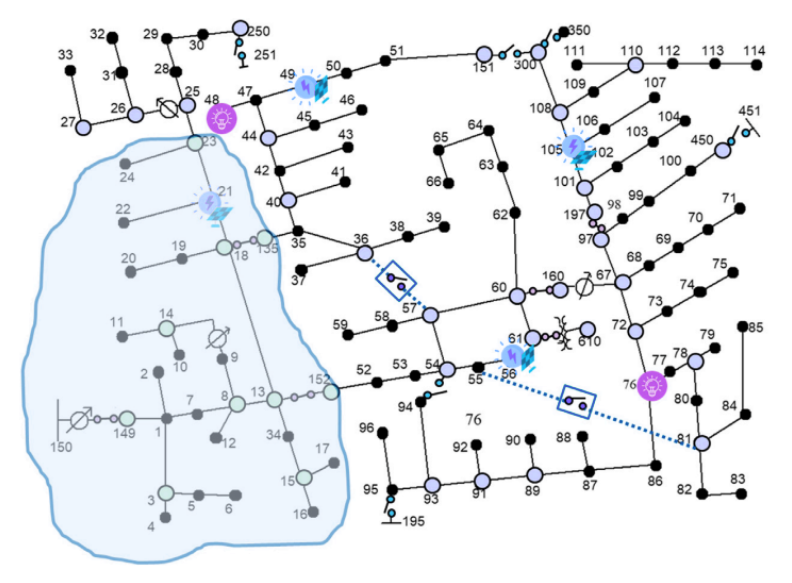}
    \caption{Flood \cite{DWIVEDI2025125001}}
    \label{ieee123}
\end{figure}

The Episode Rewards graph presents the agent’s episodic rewards over 1500 training episodes. The blue line denotes individual episode rewards, while the red line represents a 50-episode moving average. 

\begin{figure}[H]
    \centering
    \includegraphics[width=0.85\linewidth]{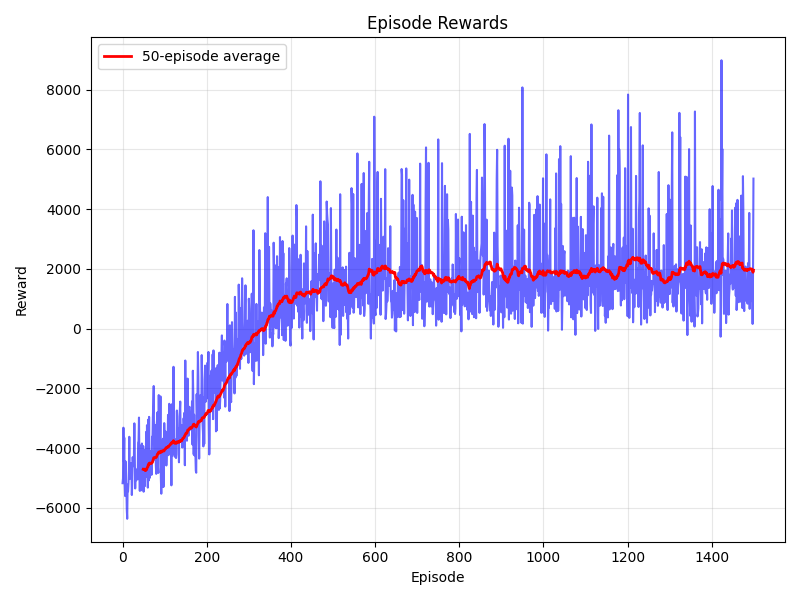}
    \caption{Reward vs Episode Graph}
    \label{C1_reward_episode_graph}
\end{figure}

During the early phase (episodes 0–400), rewards remain negative, reflecting the agent’s initial exploration and suboptimal decision-making under unfamiliar conditions. From episodes 400 to 1500, the moving average shows a consistent upward trend, indicating effective learning and policy refinement, particularly in managing grid-switching during flood-induced disruptions.


\begin{figure}[H]
    \centering
    \includegraphics[width=0.9\linewidth]{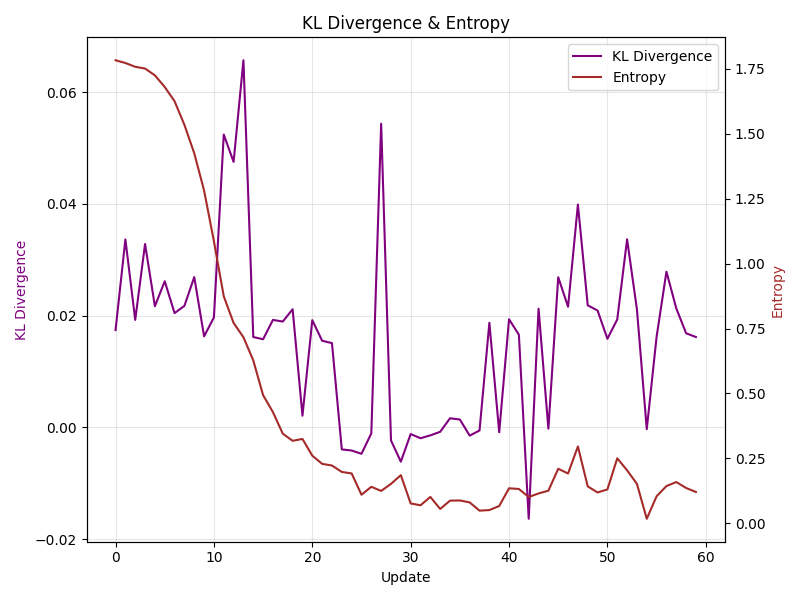}
    \caption{KL Divergence and Entropy v/s Update Graph}
    \label{C1_kl_entropy_graph}
\end{figure}

Fig.\ref{C1_kl_entropy_graph} visualizes KL divergence (purple) and entropy (red) across 60 PPO updates, providing insights into policy stability and exploration-exploitation dynamics. Initially, KL divergence values hover around 0.05, but stabilize near the target threshold of 0.01 by update 20, ensuring controlled and gradual policy improvement. Entropy, which quantifies action randomness, starts at approximately 1.75 and steadily declines to near 0.1, indicating a transition from broad exploration to focused exploitation.

\subsection{WildFire}

Consider the distribution system operating under a wildfire scenario, where certain regions (e.g., the shaded zone) are affected by network isolation.

\begin{figure}[H]
    \centering
    \includegraphics[width=0.9\linewidth]{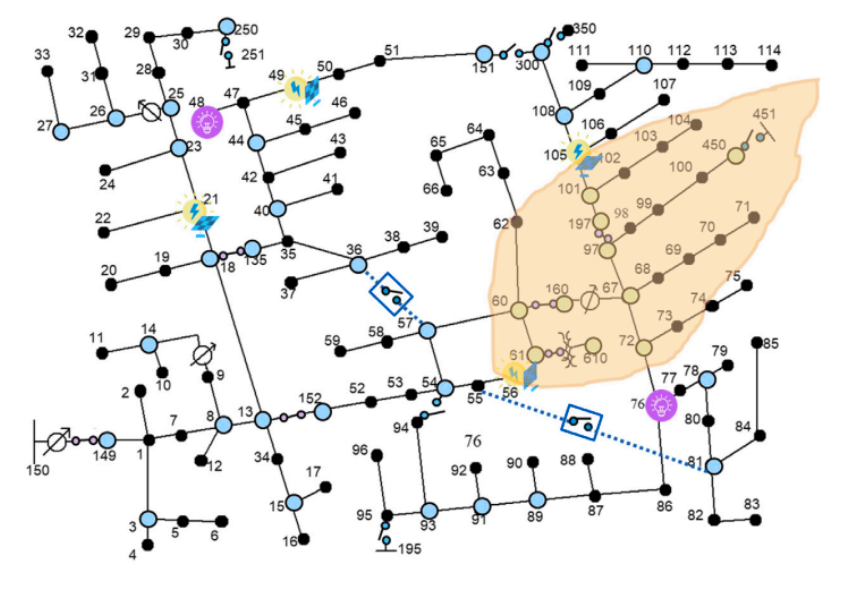}
    \caption{WildFire \cite{DWIVEDI2025125001}}
    \label{ieee123}
\end{figure}

The Episode Rewards graph shows the reward per episode (in blue) over 1400 episodes, with a 50-episode moving average (in red) to reveal the overall trend.

\begin{figure}[H]
    \centering
    \includegraphics[width=0.9\linewidth]{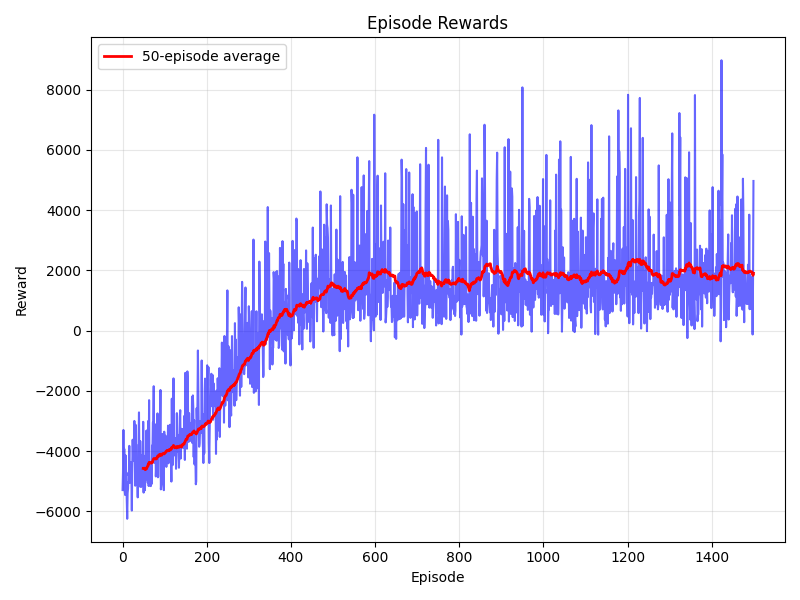}
    \caption{Reward vs Episode Graph}
    \label{C2_reward_episode_graph}
\end{figure}

Early episodes (0–200) exhibit negative rewards, indicating initial difficulties. As training continues, the average reward climbs steadily, stabilizing around 2000–4000 after episode 600, reflecting consistent improvement. Notable fluctuations, such as peaks around episodes 400 and 1000, suggest periods of enhanced performance. The leveling off post-800 episodes indicates a mature learning phase.

\begin{figure}[H]
    \centering
    \includegraphics[width=0.9\linewidth]{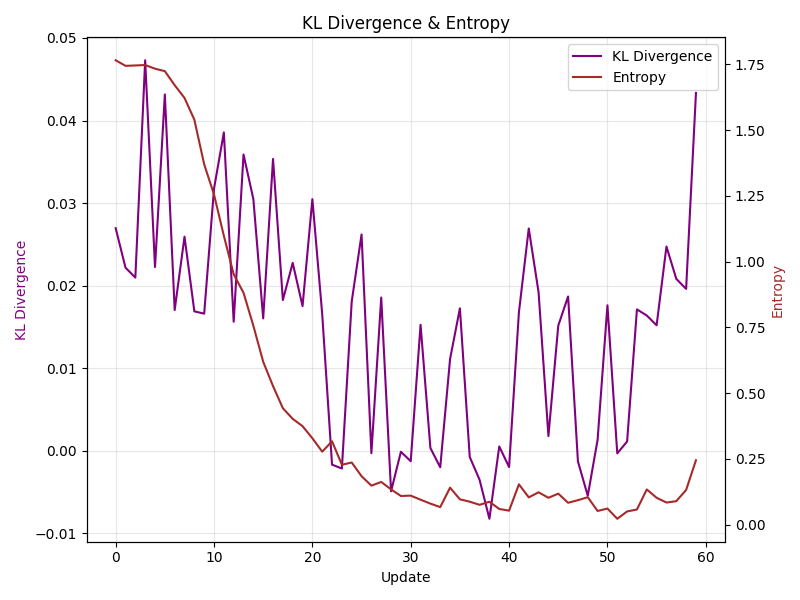}
    \caption{KL Divergence and Entropy v/s Update Graph}
    \label{C2_kl_entropy_graph}
\end{figure}

Fig.\ref{C2_kl_entropy_graph} graph plots KL divergence (purple) and entropy (red) across 60 training updates to monitor policy evolution. KL divergence starts at 0.04 and fluctuates before stabilizing near 0.01 by update 40, indicating controlled policy shifts. Entropy begins at 1.75 and declines to around 0.25, showing a move from exploration to exploitation. Spikes, such as at updates 10 and 50, highlight moments of increased variability, likely due to adaptive adjustments.



\subsection{Hurricane}

Consider the distribution system operating under a hurricane scenario, where certain regions (e.g., the shaded zone) are affected by network isolation.

\begin{figure}[H]
    \centering
    \includegraphics[width=0.9\linewidth]{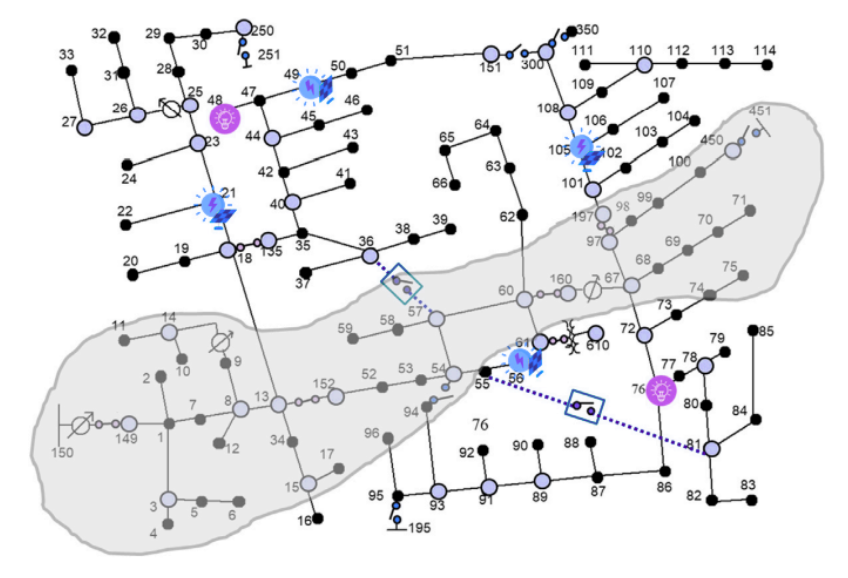}
    \caption{Hurricane \cite{DWIVEDI2025125001}}
    \label{ieee123}
\end{figure}

The Episode Rewards graph displays the reward per episode (in blue) across 1400 episodes, with a 50-episode moving average (in red) to indicate the general trend.

\begin{figure}[H]
    \centering
    \includegraphics[width=0.9\linewidth]{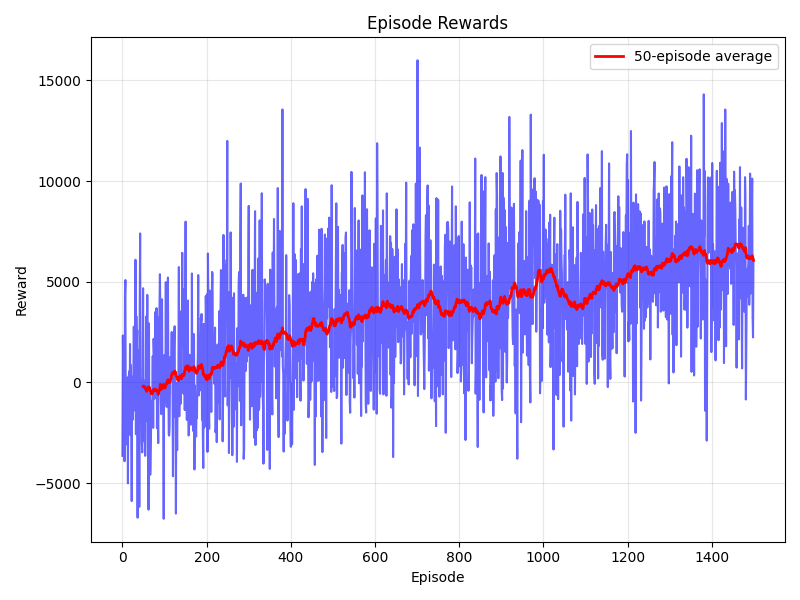}
    \caption{Reward vs Episode Graph}
    \label{C3_reward_episode_graph}
\end{figure}

 Early episodes (0–200) show negative rewards, reflecting the agent’s initial challenges. As training advances, the average reward rises progressively, peaking between 5000 and 7500 by episode 1000, showcasing improved performance in managing difficult conditions. Sharp reward spikes, such as around episodes 600 and 1000, suggest successful adaptations to high-stress scenarios. The plateau after episode 1000 indicates a stable policy. Since this case has the most destruction on the grid, the following graph underscores the agent’s resilience in recovering and maintaining performance under extreme disruption.

\begin{figure}[H]
    \centering
    \includegraphics[width=0.9\linewidth]{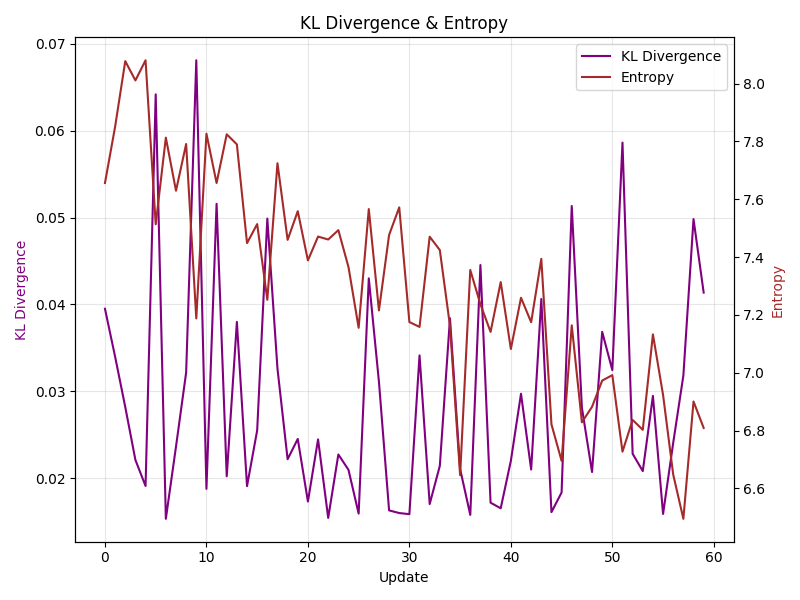}
    \caption{KL Divergence and Entropy v/s Update Graph}
    \label{C3_kl_entropy_graph}
\end{figure}

The KL Divergence \& Entropy graph tracks KL divergence (purple) and entropy (red) over 60 training updates to evaluate policy stability and exploration. KL divergence starts around 0.05 and settles near 0.01 by update 20, aligning with PPO’s target for controlled updates. Entropy decreases from 1.75 to about 0.25, indicating a transition from exploration to exploitation. Spikes in both metrics (e.g., updates 10, 25, and 35) highlight adaptive responses to environmental changes. The stabilization of both measures suggests a robust policy for dynamic grid management.

\subsection{Short-Circuit}

Consider the distribution system operating under a short-circuit scenario, where certain regions are affected by network isolation.

\begin{figure}[H]
    \centering
    \includegraphics[width=0.9\linewidth]{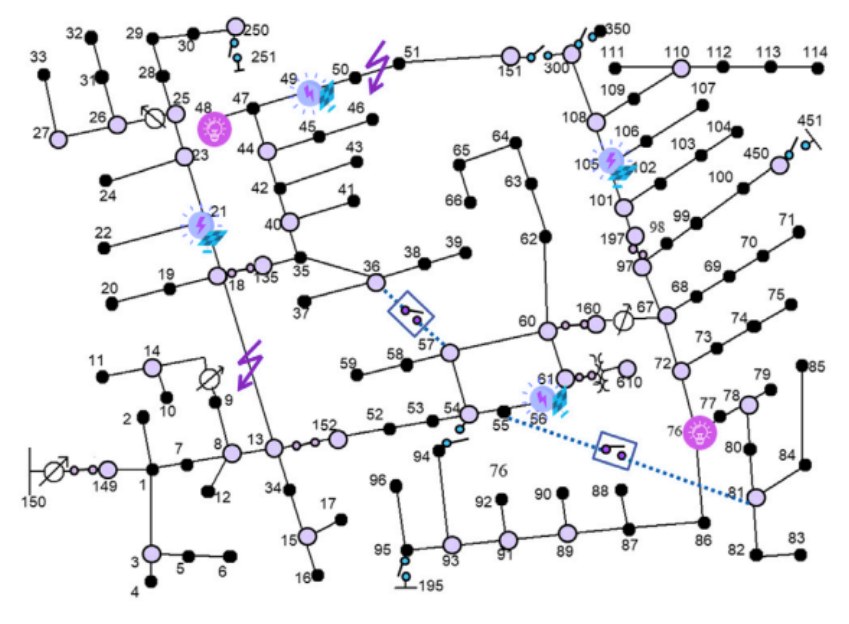}
    \caption{Short-Circuit \cite{DWIVEDI2025125001}}
    \label{ieee123}
\end{figure}


The Episode Rewards graph plots the reward obtained per episode (in blue) over 1200 episodes, with a 50-episode moving average (in red) to highlight the overall trend.

\begin{figure}[H]
    \centering
    \includegraphics[width=0.9\linewidth]{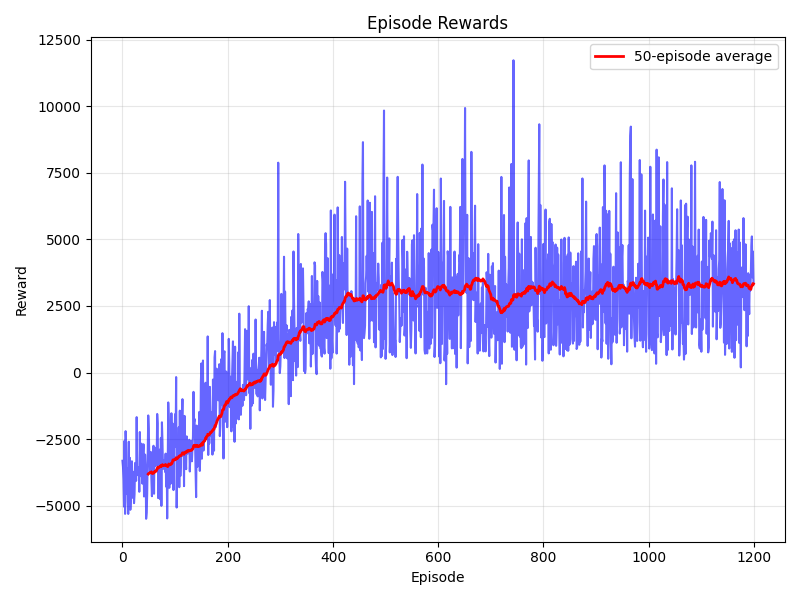}
    \caption{Reward vs Episode Graph}
    \label{C4_reward_episode_graph}
\end{figure}

Initially, rewards are negative (0–200 episodes), indicating the agent's struggle during early exploration. As training progresses, the average reward steadily increases, reaching 5000–7500 by episode 1000, demonstrating effective learning—especially in handling calamity scenarios. Occasional reward spikes (e.g., episodes 600 and 1000) reflect successful high-resilience configurations. The post-1000 plateau suggests policy convergence. The occasional peaks (e.g., around episodes 600 and 1000) likely correspond to successful adaptations to calamity conditions using high-resilience configurations (e.g., 4 DER + Additional Switches).




\begin{figure}[H]
    \centering
    \includegraphics[width=0.9\linewidth]{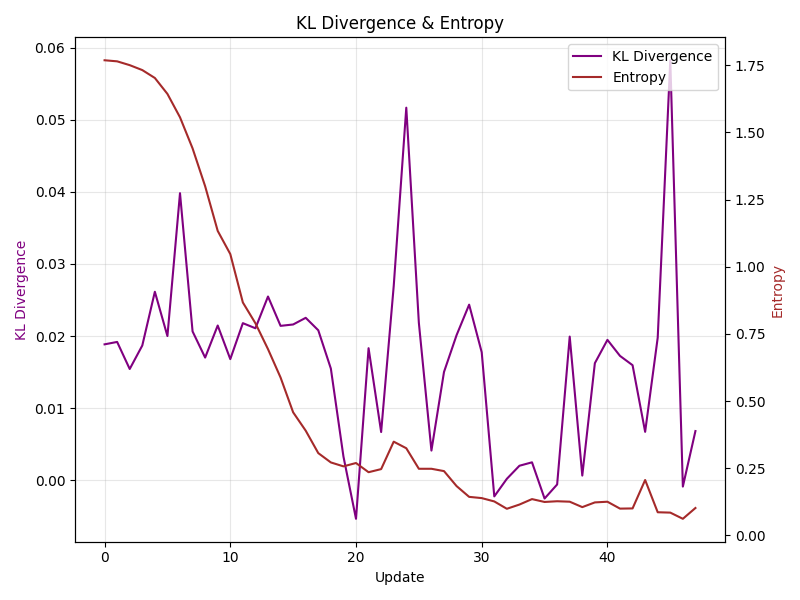}
    \caption{KL Divergence and Entropy v/s Update Graph}
    \label{C4_kl_entropy_graph}
\end{figure}

Fig.\ref{C4_kl_entropy_graph} illustrates the KL divergence (purple) and entropy (red) over 40 training updates to assess policy stability and exploration dynamics. The KL divergence, measuring deviation between successive policies, begins near 0.05 and stabilizes around the target threshold of 0.01 (as defined in PPO) by update 20, ensuring controlled policy updates. Entropy, indicative of exploration, decreases from 1.75 to approximately 0.25, reflecting a shift from exploratory to exploitative behavior. Notable spikes in both metrics (e.g., around updates 10, 25, and 35) suggest adaptive exploration in response to environmental perturbations such as simulated weather changes. The convergence of both metrics indicates the agent has stabilized to a reliable policy suitable for grid-switching under dynamic conditions.

Table \ref{commercial_analysis_table}, the Contingency Analysis Table, outlines various contingency scenarios (C1-C4) under different weather conditions (Normal, Flood, Wildfire, Hurricane, Shortcut/Additional Switch). It provides binary recommendations (0s and 1s) for each case and includes total costs in the Commercial Analysis column.

\begin{table}[h]
    \centering
    \scriptsize 
    \caption{Contingency Analysis Table}
    \label{commercial_analysis_table}
    \begin{tabular}{l l l c}
        \toprule
        Contingency & Weather & Recommendations & Commercial Analysis \\
        \midrule
        C1 & Normal & Base Case & Total Cost - \$2047654.56 \\
           &        & [1 0 1 0 0 0 1 1 0 1] & \\
           & Flood  & 3 DER & \\
           &        & [1 0 1 1 1 1 1 1 0 1] & \\
        \midrule
        C2 & Normal & Base Case & Total Cost - \$2242654.67 \\
           &        & [1 0 1 0 0 0 1 1 0 1] & \\
           & Wildfire & 3 DER & \\
           &          & [1 1 1 1 1 0 1 1 0 1] & \\
        \midrule
        C3 & Normal & Base Case & Total Cost - \$2560794.45 \\
           &        & [1 0 1 0 0 0 1 1 0 1] & \\
           & Hurricane & 2 DER & \\
           &            & [1 1 1 1 1 1 1 1 1 1] & \\
        \midrule
        C4 & Normal & Base Case & Total Cost - \$1792096.28 \\
           &        & [1 0 1 0 0 0 1 1 0 1] & \\
           & Shortcircuit & 4 DER + Additional Switch & \\
           &              & [1 1 1 1 1 1 1 1 0 1] & \\
        \bottomrule
    \end{tabular}
\end{table}




\section{Conclusion}

This research presents a novel PPO-based framework for intelligent grid switching optimization under dynamic weather conditions. The developed system successfully addresses the critical challenge of maintaining electrical grid resilience while optimizing operational costs through two complementary approaches: strategic configuration selection and tactical individual switch control.

Future research directions include extending the framework to multi-objective optimization incorporating power quality and environmental impact, integrating real-time weather forecasting and distributed energy resources, and developing transfer learning mechanisms for adaptation to new grid configurations. Emerging research frontiers such as quantum-enhanced optimization for complex grid problems, digital twin integration with real-time synchronization, and blockchain-based decentralized control mechanisms present significant opportunities for revolutionizing electrical grid management. The continued development of this PPO-based framework holds immense potential for contributing to more resilient, efficient, and sustainable power systems capable of adapting to the evolving challenges of modern electrical infrastructure and climate change.




%





\ifCLASSOPTIONcaptionsoff
  \newpage
\fi





\bibliographystyle{IEEEtran}
\bibliography{IEEEabrv,Bibliography}

\vfill


\end{document}